\documentclass[10pt,conference]{IEEEtran}
\usepackage{xspace}
\usepackage{mdframed}
\usepackage{color, colortbl}
\usepackage{array}
\usepackage{hyperref}
\usepackage{graphicx}
\usepackage{multirow}
\usepackage{soul}
\usepackage{cleveref}
\usepackage{multirow}
\usepackage{booktabs}
\usepackage{graphicx}
\usepackage{balance}
\usepackage[table,xcdraw]{xcolor}
\usepackage[switch]{lineno}

\usepackage{fancybox}
\newcommand{\hypobox}[1]{

        \begin{center}\noindent\thicklines\setlength{\fboxsep}{6pt}\cornersize{0.2}\ovalbox{

                \begin{minipage}{3.0in}

                        \textit{#1}

                \end{minipage}} 

        \end{center}} 

\mdfdefinestyle{MyFrame}{
 outerlinewidth=0pt,
 skipabove=0pt,
 skipbelow=0pt,
 innertopmargin=5pt,
 innerbottommargin=0pt,
 linewidth=0pt,
 topline=false,
 rightline=false,
 leftline=false,
 innerrightmargin=4pt,
 innerleftmargin=4pt,
 nobreak=false}

\newcommand{\GH}{{\sc GitHub}\xspace}
\newcommand{\DP}{{\sc Devpost}\xspace}
\newcommand{\WOC}{{\sc World of Code}\xspace}
\newcommand{\RQ}[2]{
\begin{mdframed}[style=MyFrame]\noindent
	\textbf{RQ}$_{#1}$.~\emph{#2}
\end{mdframed}
}

\makeatletter
\newcommand{\linebreakand}{%
  \end{@IEEEauthorhalign}
  \hfill\mbox{}\par
  \mbox{}\hfill\begin{@IEEEauthorhalign}
}
\makeatother

\definecolor{Gray}{gray}{0.9}

\newcolumntype{L}[1]{>{\raggedright\let\newline\\\arraybackslash\hspace{0pt}}m{#1}}

\usepackage{ifthen}
\usepackage{amssymb}
\DeclareOldFontCommand{\sf}{\normalfont\sffamily}{\mathsf}
\newboolean{showcomments}
\setboolean{showcomments}{true}
\ifthenelse{\boolean{showcomments}}
  {\newcommand{\nb}[2]{
    \fcolorbox{Gray}{yellow}{\bfseries\sffamily\scriptsize#1}
    {\sf\small$\blacktriangleright$\textit{#2}$\blacktriangleleft$}
   }
   
  }
  {\newcommand{\nb}[2]{}
   
  }

\begin{document}
\bstctlcite{IEEEexample:BSTcontrol}

\title{The Secret Life of Hackathon Code\\Where does it come from and where does it go?}

\author{\IEEEauthorblockN{Ahmed Imam}
\IEEEauthorblockA{University of Tartu\\
Estonia\\
ahmed.imam.mahmoud@ut.ee}
\and
\IEEEauthorblockN{Tapajit Dey}
\IEEEauthorblockA{Lero---the Irish Software Research\\
Centre, University of Limerick\\
Limerick, Ireland\\
tapajit.dey@lero.ie}
\and

\IEEEauthorblockN{Alexander Nolte}
\IEEEauthorblockA{University of Tartu, Estonia\\
Carnegie Mellon University,\\
Pittsburgh, PA, USA\\
alexander.nolte@ut.ee}
\linebreakand

\IEEEauthorblockN{Audris Mockus}
\IEEEauthorblockA{University of Tennessee\\
Knoxville, TN, USA\\
audris@utk.edu}
\and

\IEEEauthorblockN{James D. Herbsleb}
\IEEEauthorblockA{Carnegie Mellon University\\
Pittsburgh, PA, USA\\
jdh@cs.cmu.edu}}

\maketitle
\thispagestyle{plain}
\pagestyle{plain}

\begin{abstract}
Background: Hackathons have become popular events for teams to collaborate on projects and develop software prototypes. Most existing research focuses on activities during an event with limited attention to the evolution of the code brought to or created during a hackathon.
Aim: We aim to understand the evolution of hackathon-related code, specifically, how much hackathon teams rely on pre-existing code or how much new code they develop during a hackathon. Moreover, we aim to understand if and where that code gets reused, and what factors affect reuse.
Method: We collected information about 22,183 hackathon projects from \DP -- a hackathon database -- and obtained related code (blobs), authors, and project characteristics from the \WOC. We investigated if code blobs in hackathon projects were created before, during, or after an event by identifying the original blob creation date and author, and also checked if the original author was a hackathon project member. We tracked code reuse by first identifying all commits containing blobs created during an event before determining all projects that contain those commits.
Result: While only approximately 9.14\% of the code blobs are created during hackathons, this amount is still significant considering time and member constraints of such events. Approximately a third of these code blobs get reused in other projects. The number of associated technologies and the number of participants in a project increase reuse probability.
Conclusion: Our study demonstrates to what extent pre-existing code is used and new code is created during a hackathon and how much of it is reused elsewhere afterwards. Our findings help to better understand code reuse as a phenomenon and the role of hackathons in this context and can serve as a starting point for further studies in this area. 
\end{abstract}


\vspace{-5pt}
\section{Introduction}
\label{sec:intro}
Hackathons are time-bounded events during which individuals form -- often ad-hoc -- teams and engage in intensive collaboration to complete a project that is of interest to them~\cite{pe2019designing}. They have become a popular form of intense collaboration with the largest collegiate hackathon league alone reporting that their events attract more than 65,000 participants each year\footnote{\url{https://mlh.io/about}}.
The success of hackathons can at least partially be attributed to them being perceived to foster learning~\cite{porras2019code,fowler2016informal,nandi2016hackathons} and community engagement~\cite{nolte2020support,huppenkothen2018hack,taylor2018strategies,moller2014community} and tackle civic, environmental and public health issues~\cite{hope2019hackathons,taylor2018strategies,baccarne2014urban} which led to them consequently being adopted in various domains including
(higher) education~\cite{porras2019code,gama2018hackathon,kienzler2017learning},
(online) communities~\cite{huppenkothen2018hack,taylor2018everybody,busby2016closing,craddock2016brainhack}, 
entrepreneurship~\cite{cobham2017appfest2,nolte2019touched},
corporations~\cite{pe2019designing,nolte2018you,komssi2015hackathons,rosell2014unleashing}, and others. 

Most hackathon projects focus on creating a prototype that can be presented at the end of an event~\cite{medina2020what}. This prototype often takes the form of a piece of software. The creation of software code can, in fact, be considered as one of the main motivations for organizers to run a hackathon event. Scientific and open source communities, in particular, organize such events with the aim of expanding their code base~\cite{pe2019understanding,stoltzfus2017community}. 
It thus appears surprising that the evolution of the code used and developed during a hackathon has not been studied yet, as revealed by a review of existing literature.

In order to address this gap, we aim to study the evolution of the code used and created by the hackathon team members from two main perspectives. First, we study from where the code \textit{originates}: While teams will certainly develop original code during a hackathon, it can be expected that they will also utilize existing (open source) code as well as code that they might have created themselves prior to the event.

Second, to understand the impact of hackathon code, i.e. code created during a hackathon event by the hackathon team in the hackathon project repository, we aim to study whether and how it \textit{propagates} after the event has ended. There are studies on project continuation after an event has ended~\cite{nolte2020what,nolte2018you}. These studies, however, mainly focus on the continuation of a hackathon project in a corporate context~\cite{nolte2018you} and on antecedents of continuous development activity in the same repository that was utilized during the hackathon~\cite{nolte2020what}. The question of where code that has been developed during a hackathon potentially gets reused outside of the context of the original hackathon project has not been sufficiently addressed.

Moreover, we aim to understand what factors might influence hackathon code reuse, which can be useful for hackathon organizers and participants to foster the impact of the hackathon projects they organize/participate in. These factors would also be of interest to the open source community in general in order to effectively tap into the potential of hackathons as a source of new software code creation.

To cover these two perspectives, we conducted an archival analysis of the source code utilized and developed in the context of 22,183 hackathon projects that were listed in the hackathon database \DP\footnote{https://devpost.com/}. To track the origin of the code that was used and developed by each hackathon project and study its reuse after an event has ended we used the open source database \WOC~\cite{ma2019world, ma2020world} which allows us to track code usage between repositories. Overall, we looked at over 8.5M blobs~\footnote{A blob is a byte string representing a single version of a file, see  \url{https://git-scm.com/book/en/v2/Git-Internals-Git-Objects} for more details}, over 3M of which were code blobs, as identified with the help of the \GH \textit{linguist}~\footnote{\url{https://github.com/github/linguist}} tool. 

Our findings indicate that around 9.14\% of the code blobs in hackathon projects are created during an event, which is significant considering the time and team member constraints. Teams  tend to reuse a lot of existing code, primarily as in the form of packages/frameworks. Many of the projects we studied focus on front-end technologies -- JavaScript in particular -- which appears reasonable because teams often have to present prototypes at the end of an event, which lends itself to UI design. Approximately a third of code blobs created during events get reused in other projects. The number of associated technologies and the number of participants in a project increase the code reuse probability.

In summary, we make the following contributions in the paper: we present an account of code reuse both by hackathon projects and of the code generated during hackathons based on a large-scale study of 22,183 hackathon projects and 1,368,419 projects that reused the hackathon code. We tracked the origins of the code used in hackathon projects, in terms of when it was created and by whom, and also its reuse after an event. We also identified a number of project characteristics that can affect hackathon code reuse. \textbf{The replication package for our study is available at}~\cite{repPackage}.

\section{Research Questions}
\label{sec:rq}
As mentioned in~\cref{sec:intro}, the goal of this study is to understand the evolution of hackathon code and identify factors that affect code reuse for these projects. 

Our first research question thus addresses the origin of hackathon code:
\RQ{1}{Where does the code used in hackathon projects originate from?}
\vspace{-8pt}
Delving deeper into this question, we aim to understand how much of the code used in a hackathon project was actually created \textit{before} the event and reused in the project, how much of the code was developed \textit{during} the hackathon, and, since the projects sometimes continue even after the official end date of the hackathon, how much of the code was created \textit{after} the event. This leads us to the sub-question:
\RQ{1a}{When was the code created?} 
\vspace{-8pt}
We also aim to understand how much of the code in a hackathon project repository is created by one of the participants, how frequently they reused code created by someone they worked with earlier, and how much of code was created by someone else, leading us to the sub-question:
\RQ{1b}{Who were the original creators of the code?}
\vspace{-8pt}

Our second research question focuses on the aspect of hackathon code reuse. As noted in~\cref{sec:intro}, existing studies do not address the question of whether and where hackathon code gets reused after an event has ended. However, knowing the answer to this question would be crucial for understanding the impact of hackathons on the larger open source community. Some might perceive hackathons as one-off events where people gather and create some code that is never used again, while in fact they might have an impact on the wider scene of software development and create something of value that transcends individual events. Moreover, it is important to assess in which scale of project hackathon code gets reused (i.e. small projects with few developers and stars or larger projects). This aspect would be useful in understanding the impact of hackathons in greater detail, since, arguably, code that gets reused in larger projects can be perceived to have more impact on the software development community than code that is reused in smaller projects. This leads us to also asking the following second research question:
\RQ{2}{What happens to hackathon code after the event?}
\vspace{-8pt}

Finally, our third research question focuses on understanding how different characteristics of a hackathon project can influence the probability of hackathon code reuse. While code reuse in Open Source Software is a topic of much interest, there are only a few studies covering this topic. Moreover, existing studies, e.g.~\cite{haefliger2008code,kawamitsu2014identifying,feitosa2020code,von2005knowledge} only focus on between 10 and a few hundred projects. For this study, we examined 22,183 hackathon projects, which makes it reasonable to assume that insights from this study -- despite them being drawn from hackathon projects only -- would add to the existing knowledge about code reuse in general. Thus, we present our third and final research question as:
\RQ{3}{How can certain project characteristics influence hackathon code reuse?}
\vspace{-5pt}
Related to this third research question, we formed the following hypotheses that focus on aspects which can reasonably be expected to foster code reuse:\\
\noindent\textbf{H1 Familiarity:} Projects that are attempted by larger teams will have a higher chance of their code being reused, simply because more people are familiar with the code. Moreover, hackathon events that are co-located offer participants more possibilities for interaction which can contribute to a better understanding of each other's code, higher code quality, and consequently foster code reuse.\\
\noindent\textbf{H2 Prolificness:} Code from projects involving many different technologies is more likely to be reused, since: (a) they tend to have more general-purpose code than more focused projects, which affects code reuse as discussed by Mockus~\cite{mockus2007large}, and (b) they have a cross-language appeal, opening more possibility for reuse. Similarly, projects with more amount of code created before and during the event (we can not use the code created after the event, for preventing data leakage) should have a higher chance of code reuse by virtue of simply having more code.\\
\noindent\textbf{H3 Composition:} The project composition, i.e. how many blobs in a project are actually related to code, and how many are related to, e.g., data, documentation, or others could be another factor that might influence code reuse. This relationship is likely to be non-linear though, e.g., since we are considering code reuse, a higher percentage of code in the project should increase the probability of reuse, but only up to a certain point, since code from a repository containing only code and no documentation is not very likely to be reused.


\section{Background}
\label{sec:lit}
In this section we will situate our work in the context of prior research on hackathon code (\cref{sec:lit:hack}) before discussing existing studies on code reuse (\cref{sec:lit:reuse}).

\subsection{Research on hackathon code}
\label{sec:lit:hack}
The rise in popularity of hackathon event has led to an increased interest to study them~\cite{falk202010}. Current research however mainly focuses on the event itself studying how to attract participants~\cite{taylor2018everybody,hou2017hacking}, how to engage diverse audiences~\cite{paganini2020engaging,hope2019hackathons,filippova2017diversity}, how to integrate newcomers~\cite{nolte2020support}, how teams self-organize~\cite{trainer2016hackathon} and how to run hackathons in specific contexts~\cite{moller2014community,pe2019designing,porras2019code}. These studies acknowledge the project that teams work on as an important aspect. The question of where the software code that teams utilize for their project comes from and where it potentially gets reused after an event has not been a strong focus though.

There are also studies that focus on the continuation of software projects after an event has ended~\cite{lapp20072006,cobham2017appfest2,nolte2018you,ciaghi2016hacking}. These studies however mainly discuss how activities of a team during, before, and after a hackathon can foster project continuation~\cite{nolte2018you}, how hackathon projects fit to existing projects~\cite{lapp20072006}, and the influence of involving stakeholders when planning a hackathon project on its continuation~\cite{cobham2017appfest2,ciaghi2016hacking}. They do not specifically focus on the code that is being developed as part of a hackathon project.

Few studies have also considered the code that teams develop during a hackathon~\cite{nolte2020what,busby2016closing}. These studies however mainly focus on code availability after an event~\cite{busby2016closing} or on how activity before and after an event within the same repository that a team utilized during the hackathon can affect reuse~\cite{nolte2020what}. The question of whether and to what extend teams utilize existing code and whether and where the code that they develop during a hackathon gets reused aside from this specific repository has not been addressed.

\subsection{Code reuse}
\label{sec:lit:reuse}
Code reuse has been a topic of interest and is generally perceived to foster developer effectiveness, efficiency, and reduce development costs~\cite{sojer2010code,haefliger2008code,feitosa2020code}. Existing work so far mainly focuses on the relationship between certain developer traits~\cite{haefliger2008code,sojer2010code,von2005knowledge} and team and project characteristics such as team size, developer experience, and project size and code reuse~\cite{abdalkareem2017code}. Moreover, the aforementioned findings are mainly based on surveys among developers, thus covering their perception rather than actual reuse behavior. In contrast, we aim to study actual code reuse behavior.

There is also existing work that focuses on studying the reuse of the code itself. These, however, are often small scale studies of a  few projects~\cite{kawamitsu2014identifying,xu2020reinventing} focusing on aspects such as automatically tracking reuse between two projects~\cite{kawamitsu2014identifying} and identifying reasons why developers might choose reuse over re-implementation~\cite{xu2020reinventing}. In contrast, our aim is to study how the code created during a hackathon evolves i.e. where it comes from and whether and where it gets reused.

Large scale studies on code reuse have been scarce. The few existing studies often focus on code dependencies~\cite{german2007using} or on technical dept induced based on reuse~\cite{feitosa2020code} which are both not a strong focus for us because our aim is rather to study where hackathon code gets reused. There are studies that discuss the reuse of code on a larger scale~\cite{mockus2007large} and showed that it is mainly code from large established open source projects that get reused, while we aim to study reuse of code that has been developed by a small group of people during a short-term intensive coding event. 

\section{Methodology}
\label{sec:method}

\vspace{-10pt}
\subsection{Data Sources}
While hackathon events have risen in popularity in the recent past, many of them remain ad-hoc events, and thus data about those events is not stored in an organized fashion. 
However, \DP is a popular hackathon database that is used by corporations, universities, civic engagement groups and others to advertise events and attract participants. It contains data about hackathons including hackathon locations, dates, prizes and information about teams and their projects including the project's \GH repositories. Organizers curate the information about hackathons and participants indicate which hackathons they participated in, which teams they were part of and which projects they worked on. \DP does not conduct accuracy checks. 

However, \DP does not contain all the information required for answering our research questions. We thus leveraged the \WOC dataset for gathering additional information about projects, authors, and code blobs. \WOC is a prototype of an updatable and expandable infrastructure to support research and tools that rely on version control data from the entirety of open source projects that use Git. 
It contains information about OSS projects, developers (authors), commits, code blobs, file names, and more. \WOC provides maps of the relationships between these entities, which is useful in gathering all relevant information required for this study.
We used version S of the dataset for the analysis described in this paper which contains repositories identified until Aug 28, 2020.

\vspace{-5pt}
\subsection{Data Collection and Cleaning}
\begin{figure}[t]
    \centering
    \includegraphics[width=\linewidth]{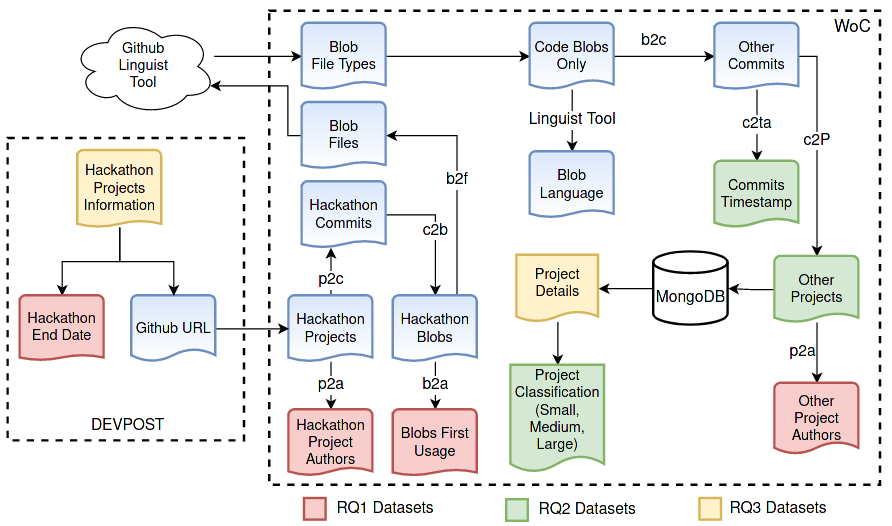}
    \caption{\textbf{Data Collection Workflow: Highlighting the different data sources used and the process of gathering the required information from them, and the data used in answering our research questions}}
    \label{fig:DataCollection}
    \vspace{-15pt}
\end{figure}

Here we describe how we collected the data required for answering our research questions, along with details of all the filtering we introduced. An overview of the approach is shown in~\cref{fig:DataCollection}, which also highlights the different data sources and what data was used for answering each research question.

\subsubsection{Selecting appropriate hackathon projects for the study}
We started by collecting information about 60,479 hackathon projects from \DP. Since the project ID used in \DP is different from the project names in \WOC, in order to link these hackathon projects to the corresponding projects in \WOC, we looked at the corresponding \GH URLs, which could be easily mapped to the project names used in \WOC, where the project names are stored as \texttt{GitHubUserName\_RepoName}. After filtering out the projects without a \GH URL, we ended up with 23,522 projects. While trying to match these projects with the corresponding ones in \WOC, we were not able to match 1,339 projects, which might have been deleted or had their names changed afterwards. Thus, we ended up with 22,183 projects for further analysis.

\subsubsection{Gathering the contents (blobs) of the project}
Our first step was to identify all code blobs used in the hackathon projects. \WOC does not have a direct map between projects and blobs, so we started by collecting commits for all hackathon projects using the project-to-commit (\textit{p2c}) map in \WOC, which covers all commits for each hackathon project.
For the 22,183 hackathon projects, we collected 1,659,435 commits generated in the hackathon repository. 
Then, we gathered all the blobs associated with these commits using the commit-to-blob (\textit{c2b}) map in \WOC, which yielded 8,501,735 blobs, which are all the blobs associated with the hackathon projects under consideration.

\subsubsection{Filtering to only select code blobs}
The hackathon project repositories, like most other OSS project repositories, have more than just code in them --- they also contain images, data, documentation, etc. Since our aim in this project is the identification of the reuse of ``code", we decided to filter the blobs to only have the ones related to ``code". In order to achieve that, we looked at the filenames for each of the blobs in the project (since blobs only store the contents of a file, not the file name) using the blob-to-filename (\textit{b2f}) maps in \WOC. After that, we used the \textit{linguist} tool from \GH to find out the file types. The \textit{linguist} tool classifies files into types ``data, programming, markup, prose'', with files of type ``programming'' being what we are focusing on in this study. Additionally, we marked all the files that are not classified by the tool as files of type ``Other'', with the presumption that they do not contain any code. Therefore, we focused only on the blobs whose corresponding files are classified as ``programming'' files by the \textit{linguist} tool, which reduced the number of blobs under consideration to 3,079,487.

\begin{table*}[!htb]
\caption{Description of variables used for addressing RQ3. For the Binary variable, no. of TRUE/FALSE cases are shown}
\label{t:variables}
\resizebox{\textwidth}{!}{%
\begin{tabular}{@{}llp{5cm}p{2.5cm}rr@{}}
\toprule
\textbf{Hypothesis} & \textbf{Variable} & \textbf{Variable Description} & \textbf{Source} & \textbf{Value Range (min-max)} & \textbf{Median} \\ \midrule
 & no.Participant & Number of Participants in the hackathon & \DP & 2 - 10 & 3 \\
\multirow{-2}{*}{H1: Familiarity} & is.colocated & hackathon is held in single or multi location & \DP & \multicolumn{2}{l}{\cellcolor[HTML]{FFFC9E}TRUE: 12,445 (97\%) FALSE:436 (3\%)} \\\hline
 & no.Technology & Number of different technologies the hackathon is related to & \DP & 1 - 40 & 5 \\
 & Before & Number of blobs in the hackathon project repo that were created before the event & \WOC & 0 - 205,666 & 4 \\
\multirow{-3}{*}{H2: Prolificness} & During & Number of blobs in the hackathon project repo that were created during the event & \WOC & 1 - 3288 & 23 \\\hline
 & pctCode & Fraction of the blobs in the project repo that are classified as ``programming'' & \WOC and \GH & 0 - 1 & 0.40 \\
 & pctMarkup & Fraction of the blobs in the project repo that are classified as ``Markup'' & \WOC and \GH & 0 - 0.96 & 0.02 \\
 & pctData & Fraction of the blobs in the project repo that are classified as ``Data'' & \WOC and \GH & 0 - 0.999 & 0.12 \\
\multirow{-4}{*}{H3: Composition} & pctProse & Fraction of the blobs in the project repo that are classified as ``Prose'' & \WOC and \GH & 0 - 0.999 & 0.03 \\ 
\bottomrule
\end{tabular}%
}
\vspace{-15pt}
\end{table*}

\subsubsection{Gathering data required to identify the origins of hackathon code (RQ1)}
\label{sss:rq1}
To address our first research question, we needed information about the first commits associated with each of the 3,079,487 blobs under consideration. Fortunately, it is possible to get information about the first commit that introduced each blob using \WOC. We extracted the \textit{author} of that first commit, along with the \textit{timestamp}, which would be useful in identifying when the blob was first created. We have the end date for each of the hackathon events from \DP, however, it does not include any information about the start date of the hackathon. We consider the start of a hackathon 72 hours before the end date. This assumption appears reasonable since hackathons are commonly hosted over a period of 48 which are often distributed over three days~\cite{Nolte2020HowTO,cobham2017appfest2}. We also conducted a manual investigation of 73 randomly selected hackathons and found that only 2 projects (2.7\%) were longer than 3 days, which empirically suggested that our assumption would be valid for most of the hackathons under consideration.
Under this assumption, we have the start and end dates of each of the hackathon events, and we used that information to identify if a blob used in a hackathon project was created before, during, or after the hackathon event.

We can identify the first commit that introduced the hackathon blobs under consideration, the \textit{author} of that commit, and all of the developers who have been a part of the hackathon project~\footnote{We used the approach outlined by~\cite{fry2020idres} for author ID disambiguation to merge all of the different IDs belonging to one developer together, which is a common occurrence, as discussed in~\cite{Dey2020RepresentationOD}} using \WOC. With this data, we can determine if the blob was first created by a member of the hackathon team or someone else. In order to dig further and understand if the blob was created in another project a member of the hackathon project participated in, we used \WOC to identify the project associated with the first commit for each blob under consideration, identified all developers of that project, and checked if any of them are members of the team that created the hackathon project under consideration. This lets us identify if the blob was created by (a) a developer who is a participant of the hackathon project (thus, they are creating the code/reusing what they had created earlier), (b) a developer who was part of a project one of the participants of the hackathon project also contributed to (which might suggest that they are familiar with the code, which might have influenced the reuse of that code in the hackathon project), or (c) someone else who has not contributed to any of the projects the hackathon project developers previously contributed to (which would suggest a lack of direct familiarity with the code from the hackathon participants' perspective).

\subsubsection{Gathering data to identify hackathon code reuse (RQ2)}
\label{sss:rq2}
Our second research question focuses on the reuse of \textit{hackathon code}, which, per our definition (see ~\cref{sec:intro}), refers to the blobs created during the hackathon event by one of the members of the hackathon team. Therefore, to address this question, we utilized the results of our earlier analysis in order to focus only on the code blobs which satisfy the following two conditions: (a) The blob was first introduced during the hackathon event and (b) the blob was created by one of the hackathon project developers. After identifying 581,579 blobs that met these conditions, we collected all commits containing these blobs from \WOC using the blob-to-commit (\textit{b2c}) map, and we collected the projects where these commits are used using the commit-to-project (\textit{c2p}) map. \WOC has the option of returning only the most central repositories associated to each commit, excluding the forked ones (based on the work published in~\cite{forkrepo}), and we used that feature to focus only on the repositories that first introduced these blobs, and excluded the ones that were forked off of that repository later, since most forks are created just to submit a pull request and counting such forks would lead to double-counting of code reuse.


In addition to understanding how the blobs get reused, we also wanted to understand if they are reused in very small projects, or if larger projects also reuse these blobs. So, we needed a way to classify the projects into different categories. We focused on two different project characteristics for the purpose of such classification: the number of developers who contributed to that project, and the number of \textit{stars} it has on \GH, a measure available from a database (MongoDB) associated with \WOC. 
Both the number of developers and \textit{stars} are quintessential measures of project size and popularity and were found to have a low correlation (Spearman Correlation: 0.26), so we decided to use both measures.
Instead of manually classifying the projects using these variables using arbitrary thresholds, we decided to use \textit{Hartemink's pairwise mutual information based discretization method}~\cite{hartemink2001principled}, which was applied to a dataset with log-transformed values of the number of stars and developers for projects, to classify them into three categories: Small, Medium, and Large. We found different thresholds for the number of developers and \textit{stars} (for no. of developers, $>2 \rightarrow$ Medium projects and $>6 \rightarrow$ Large; for stars, $>1 \rightarrow$ Medium and $>14 \rightarrow$ Large), and classified a project as ``Large" if it is classified as such by either the number of developers or the number of \textit{stars}, and used a similar approach for classifying them as ``Medium''. The remaining projects were classified as ``Small''. Overall, we identified 1,368,419 projects that reused at least one of the 581,579 blobs, and using our classification, 1,220,114 (89.2\%) projects were classified as ``Small'', 116,177 (8.5\%) as ``Medium'', and 32,128 (2.3\%) as ``Large''.

\subsubsection{Collecting Data for Identifying the factors that affect hackathon code reuse (RQ3)}
\label{sss:rq3-data}
In addition to tracking hackathon code reuse (RQ2), we also aimed to study factors that can affect this phenomenon. For this purpose we collected various characteristics of the hackathon projects, both from \DP and \WOC, and extracted the variables of interest, per the hypotheses presented in \cref{sec:rq}.

The data we collected for RQ2 was for the blobs, so, in order to find out if code from a project were reused, we investigated how many blobs from a project was reused, and calculated the ratio of the number of reused code blobs and the total number of code blobs in the project. This revealed that almost 60\% of projects had none of their code reused. So, we decided to pursue a binary classification problem for predicting if a project has at least one code blob reused or not instead of doing regression analysis.

For the purpose of our analysis, we excluded hackathon projects with a single member, since a hackathon project ``team'' with a single participant does not really make a lot of sense, and also the projects that were not related to any existing technology, since these likely were non-technical events. By looking for code reuse, we also automatically filtered out any project that had no code blobs in its repository. After these filterings, we were left with 12,881 hackathon projects.

For the variables related to \textbf{H3}, the composition of the repository, we have 5 categories, 4 of which are dictated by the \GH \textit{linguist} tool: \textit{Code(programming), Markup, Data}, and \textit{Prose}, and a category \textit{Other} for all file types not classified by the tool. We looked at what percentage of the blobs in the projects belonged to which type. Since they all sum up to 100\%, 4 of these variables are sufficient to describe the fifth variable. In order to remove the resulting redundancy, we decided to remove the entry for type \textit{Other}, since its effect is sufficiently described by the remaining variables.

The description of all the variables along their sources and values are presented in \cref{t:variables}.

\subsubsection{Analysis Method for Identifying project characteristics that affect code reuse (RQ3)}
\label{sss:rq3-analysis}
As we noted in the hypotheses presented in \cref{sec:rq}, we are expecting some of the project characteristics to have a linear effect on hackathon code reuse, while some should have a more complex non-linear effect. The goal of our analysis is not to make the best predictive model that gives the optimum predictive accuracy, instead, we are trying to find out which of the predictors have a significant effect by creating an explanatory model. As noted by Shmueli~\cite{shmueli2010explain}, these two are very different tasks.

In order to achieve our goal of having linear and non-linear predictors in the same model and be able to infer the significance of each of them, we decided to use Generalized Additive Models (GAM). Specifically, we used the implementation of GAM from the \texttt{mgcv} package in \textit{R}. 


\section{Results}
\label{sec:results}

Here we will discuss our findings in relation to our research questions and discuss the result of a small case-study on some examples of code reuse for selected hackathon projects.

\subsection{Origins of hackathon code (RQ1)}
\label{sec:results:rq1}

\begin{figure}[t]
\centering
\vspace{-10pt}
\includegraphics[width=0.6\linewidth]{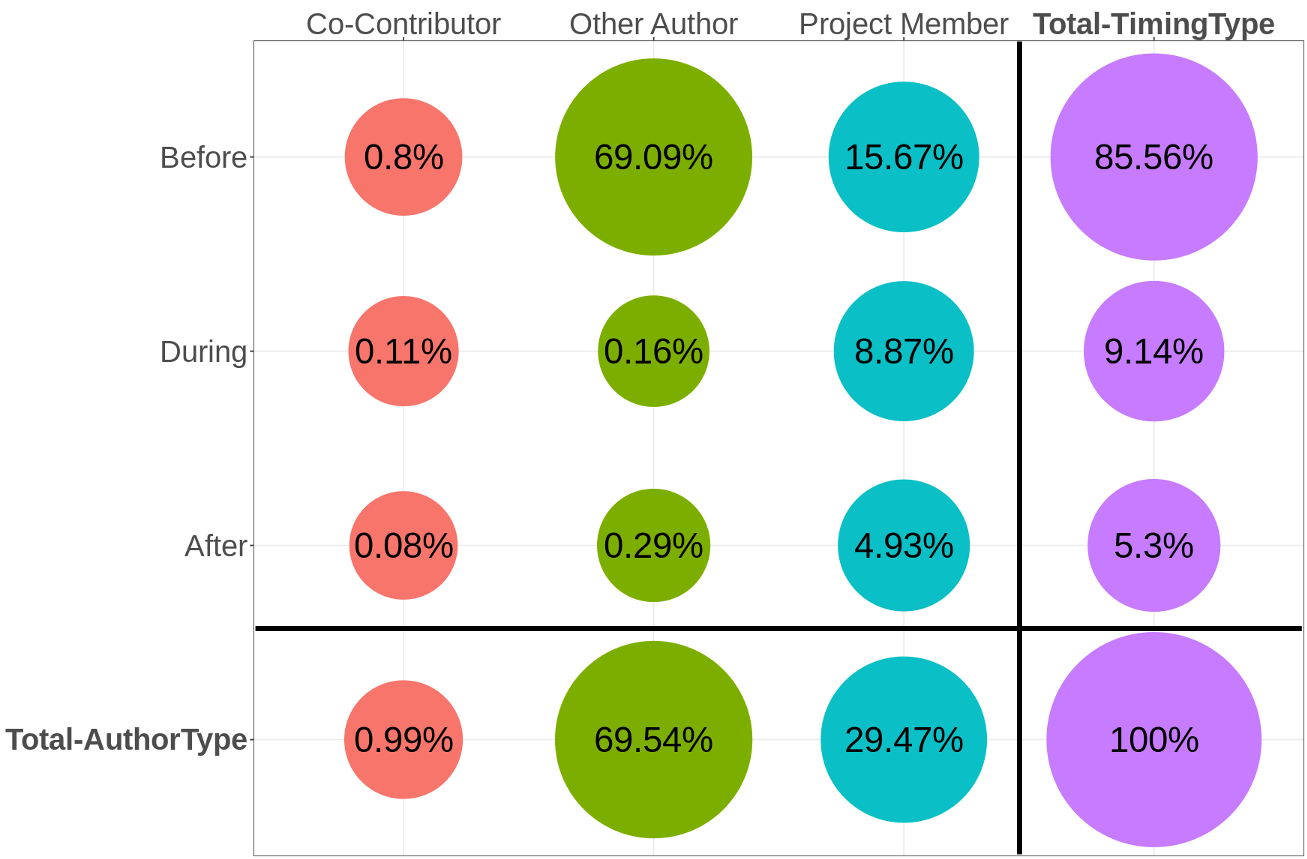}%
\vspace{-10pt}
\caption{\textbf{Plot of Who created how much of the Hackathon Code and When}}
\label{fig:rq1}
\vspace{-15pt}
\end{figure}

As mentioned in \cref{sec:rq} and \cref{sss:rq1}, we focused on two aspects while looking for the origins of the code in the hackathon project repositories, when was it created (RQ1.a), and who was the original creator of the code blob (RQ1.b).
In terms of ``when'', we examined if the first creation of the code blob under consideration was \textit{Before, During}, or \textit{After} the corresponding hackathon event. In terms of ``who'', we checked if the first creator of the code blob was one of the members of the hackathon project (\textit{Project Member}), or someone who was a contributor to a project in which one of the members of the hackathon project contributed to as well (\textit{Co-Contributor}), or someone else (\textit{Other Author}).

The result of the analysis is presented in \cref{fig:rq1}, showing that, overall, 85.56\% of the code used in the hackathon projects was created before an event. Most of these reused blobs were part of a framework/library/package used in the hackathon project, which aligns with the findings of~\cite{mockus2007large}. Around 9.14\% of the blobs were created during events, since participants need to be efficient during an event owing to the time limit~\cite{nolte2018you} which fosters reuse as previously discussed in the context of OSS~\cite{haefliger2008code}. We also found that 5.3\% of the blobs were created after an event, suggesting that most teams do not add a lot of new content to their hackathon project repositories after the event. This finding is in line with prior work on hackathon project continuation~\cite{nolte2020what}. 

\begin{figure}[!t]
\centering
\vspace{-10pt}
\includegraphics[width=0.9\linewidth]{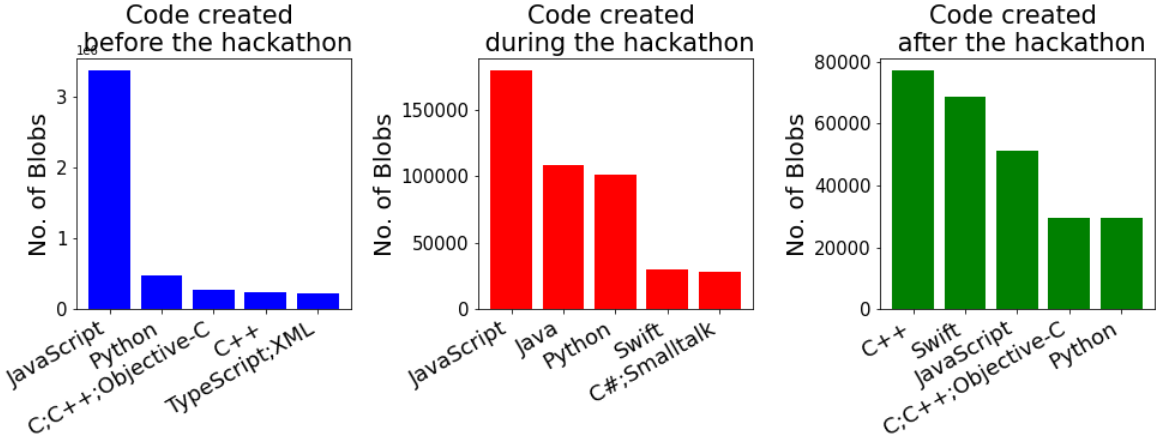}%
\vspace{-10pt}
\caption{\textbf{Top 5 languages for blobs created before, during, and after hackathons}}
\label{fig:rq1-time}
\vspace{-15pt}
\end{figure}

Looking at top languages for the blobs created at different times (\cref{fig:rq1-time}), identified by the \GH \textit{linguist} tool, we found that most of the code reused by hackathon projects (created before) is  JavaScript, and other top languages together indicate that most of the reused code by hackathon projects are related to web development frameworks. JavaScript is also the most common language worked on during the hackathons we studied, followed by Java, Python, Swift, and C\#/Smalltalk, indicating that most hackathon projects work on developing web/mobile apps. C++ was the most common language for code developed after the event, followed by Swift and JavaScript, showing a slight shift in the type of work done after the event, favoring Machine Learning applications. 

\hypobox{\textbf{RQ1.a}: 85.56\% of the code (in terms of the no. of blobs) in the hackathon project repositories is created before the hackathons, with around 9.14\% of the code being created during the events (which is significant considering the limited duration of the hackathons).} 

\begin{figure}[!t]
\centering
\includegraphics[width=0.9\linewidth]{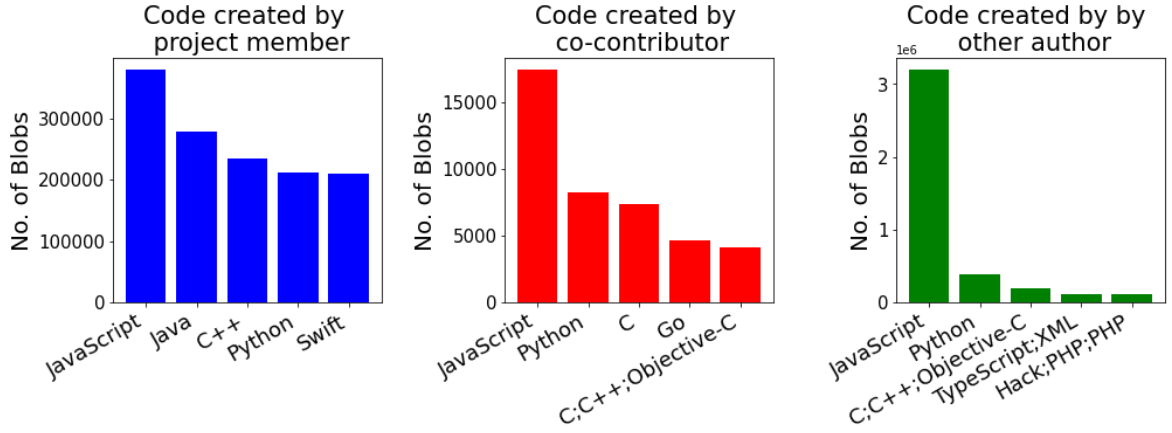}%
\vspace{-10pt}
\caption{\textbf{Top 5 languages for blobs created by project members, co-contributors, and others}}
\label{fig:rq1-author}
\vspace{-15pt}
\end{figure}

Figure \ref{fig:rq1} shows that, overall, the original creators of most of the code blobs (69.54\%) in the hackathon project repositories are someone who is not a part of the team. They are mostly the original creators of some project/package/framework used by the hackathon team. Around one-third (29.47\%) of the code was created by the project members, and the reuse of code from co-contributors in other projects is very limited (0.99\%). This aspect has not been extensively studied in the context of work on hackathons yet.

Looking at the top languages for the code created by different authors, as shown in \cref{fig:rq1-author}, we can see that, once again, most of the code created by developers not part of the hackathon team is JavaScript, which is similar to the code created before the event (\cref{fig:rq1-time}). This is not surprising, since they have a great deal of overlap (\cref{fig:rq1}). Most of the code created by project members indicate a leaning towards web/mobile app development, and most of the C++ and Python code was found to be related to Machine Learning frameworks. 

\hypobox{\textbf{RQ1.b}: The members of the hackathon teams created around 29.47\% of the code blobs, while 69.54\% of the code blobs are created by developers outside the team (mostly authors of some project/package/framework used by the team).}

If we consider the code blobs created during and after the hackathon event, as shown by the combined picture in \cref{fig:rq1}, which are the main contributions of the hackathons in terms of code creation, we see that most of that code (97\% for the code created during the event, and 93\% of the code created after the event) was actually created by the project members. Moreover, we also find that the project members often reuse the code they had written earlier in their projects, 15.67\% of all the code belong to this category. This finding is in line with prior work on hackathon projects in that teams often prepare their projects e.g. by setting up a repository~\cite{nolte2020what} and/or making (detailed) plans on what they want to achieve during an event~\cite{nolte2018you}. 

It is also worth noting that, typically, widely-used frameworks like Django, Rails, jQuery, etc have decades of development history and a large number of contributors. Thus, around 9.14\% of blobs being created during the short duration of hackathon events (72 hours per our assumption) by teams of around 2-3 members (up to a maximum of 10 members --- see \cref{t:variables}), is indeed significant and it highlights the importance of hackathons in generating new code.

\hypobox{\textbf{Origin of the Hackathon Code (RQ1)}: Hackathon projects often reuse code in terms of some package/framework. Teams also tend to reuse their own code. Most of the code created during or after the event is created by the hackathon team members.}

\subsection{Hackathon code reuse}
\label{sec:results:rq2}


\begin{figure}[t]
\centering
\vspace{-10pt}
\includegraphics[width=0.9\linewidth]{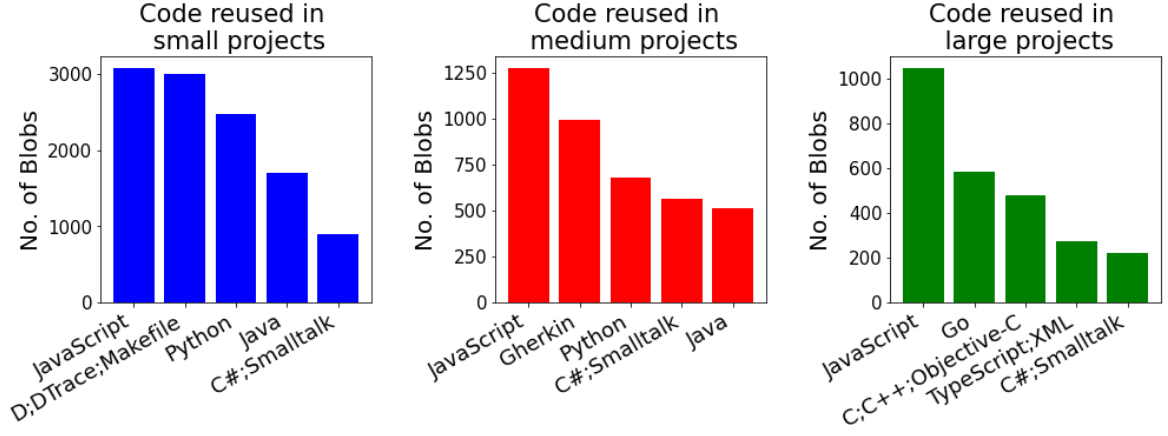}%
\vspace{-10pt}
\caption{\textbf{Top 5 Languages for the reused code blocks in different projects}}
\label{fig:rq2-lang}
\vspace{-15pt}
\end{figure}

As discussed in \cref{sec:rq} and \cref{sss:rq2}, our goal while looking for hackathon code reuse is twofold: First, we want to see how much of the code gets reused, and second, we want to find if they get reused in small, medium, or large projects. By following the procedure outlined in \cref{sss:rq2}, we found that 167,781 (28.8\%) of the 581,579 hackathon code blobs got reused in other projects. 

We further classified the projects that reused these code into \textit{Small, Medium}, and \textit{Large}, as discussed in \cref{sss:rq2}. To recap, 89.2\% of the projects that reused the hackathon code blobs were classified as \textit{Small}, 8.5\% were \textit{Medium}, and 2.3\% were classified as \textit{Large} projects. By investigating the blobs reused by these projects we found that, unsurprisingly, there are a number of instances where a blob was reused in projects of different categories.  However, such cases were found to be quite rare, in fact, only 8.85\% of reused blobs got reused in more than one project. By looking at the size of the projects a blob was reused in, we found that over half (57.73\%) of the blobs are only reused in other \textit{Small} projects, around one-third (32.85\%) are reused in \textit{Medium} projects, and less than a tenth (9.42\%) are reused in \textit{Large} projects. 

The top-5 languages for the blobs reused by various projects are shown in \cref{fig:rq2-lang}. As we can see JavaScript still remains the most common, and Python, C/C++, C\#/Smalltalk, Java were among the top ones as well. While most reused blobs are related to web/mobile apps/frameworks, we also found the relatively uncommon Gherkin being the second most common language for \textit{Medium}  projects, and the \textit{Small} projects reused a lot of blobs related to D/DTrace/Makefile. 

\begin{figure}[!t]
\centering
\vspace{-10pt}
\includegraphics[width=0.8\linewidth]{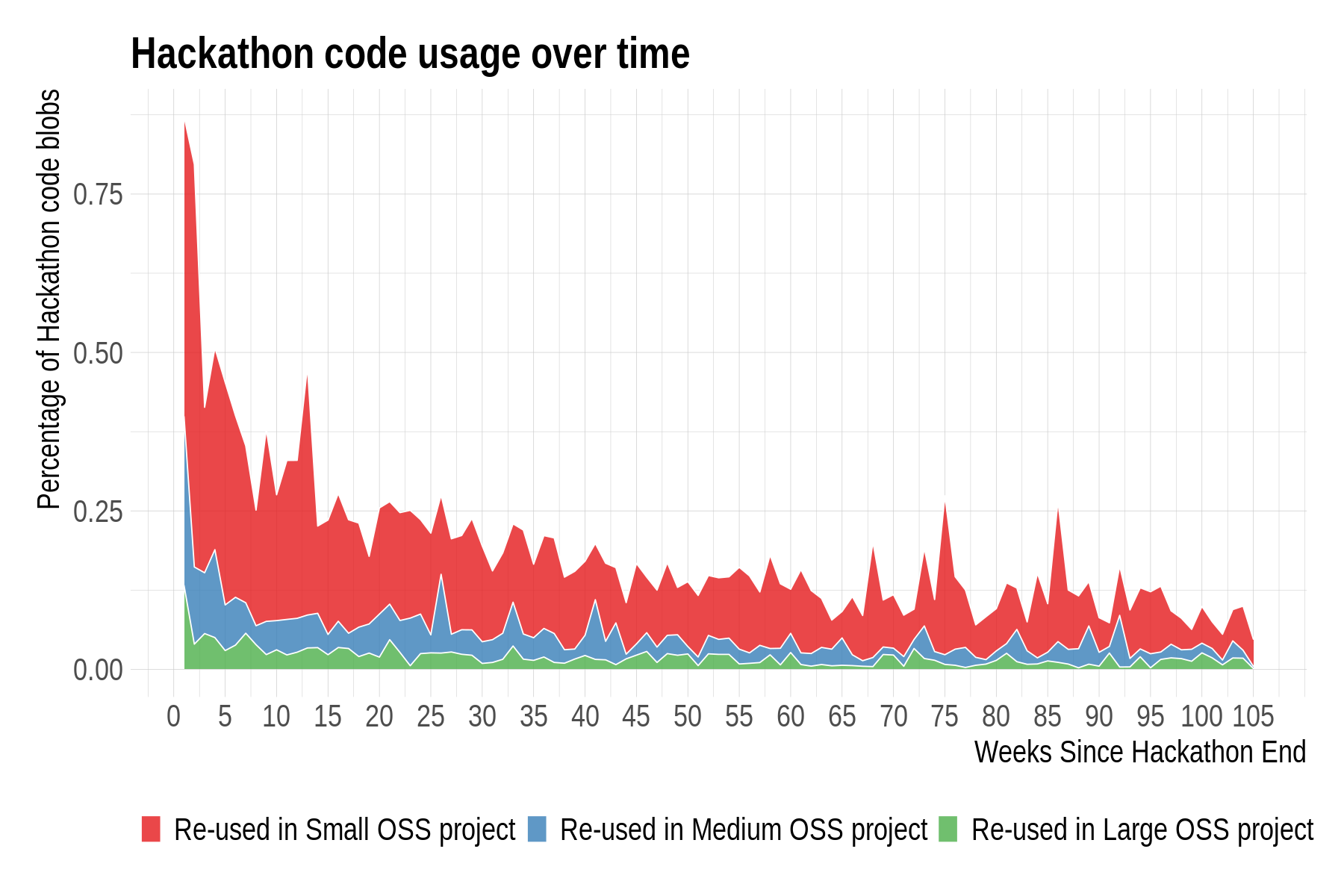}%
\vspace{-10pt}
\caption{\textbf{Plot depicting hackathon weekly code reuse in projects of different categories over the period of 2 Years}}
\label{fig:rq2-time}
\vspace{-15pt}
\end{figure}

We were interested in exploring the temporal dynamics of code reuse as well. Therefore, we looked at the reuse of hackathon code blobs over-time for the duration of two years (104.3 weeks) after the corresponding hackathon event ended. The result of that analysis is shown in \cref{fig:rq2-time}, which shows the \textit{weekly} hackathon code reuse for 2 years after the end of the corresponding hackathon event, with the fraction of total number of hackathon code blobs (581,579) reused per week on the Y-axis. 
As we can see from this plot, while overall 28.8\% of the hackathon code blobs were reused, over the span of a single week, no more than 0.8\% of the blobs got reused. This finding is in line with prior work on hackathon project continuation (e.g. Nolte et al.~\cite{nolte2020what} found that continuation activity drops quickly within one week after a hackathon before reaching a stable state) within the same repository that the team used during the hackathons. A clear trend of the code reuse dropping and then saturating after some time is visible, which is significant because it indicates that the \textit{code created in the  hackathon events continue to bear some value even after 2 years} have passed after the event. For code reuse in \textit{Small} projects, the knee point comes after around 10-15 weeks, while for the \textit{Medium} and \textit{Large} projects, it comes much earlier, in around a month. It is also a bit surprising to see code reuse peaking so soon after the event, but this could be due to the participants of the event putting/influencing people they know to put the code they think is valuable to some other project where they think it might be of use. This distinction has not been studied in prior work on hackathon code.

\hypobox{\textbf{Hackathon code reuse (RQ2):} Around 28.8\% of hackathon code blobs got reused in other projects, with 57.73\% of the code being used in \textit{Small} projects, 32.85\% in \textit{Medium} projects, and 9.42\% in \textit{Large} projects. Most of the reused blobs were related to web/mobile apps/frameworks. The temporal dynamics of code reuse show a clear trend of it reducing over time, and then saturating to a stable value.}

\subsection{Characteristics affecting code reuse}
\label{sec:results:rq3}

Our third and final Research Question was about identifying what project characteristics affect code reuse and we formed three hypotheses about what factors might be affecting it, which were presented in \cref{sec:rq}. Using the procedure outlined in \cref{sss:rq3-data}, we gathered the variables of interest related to the three hypotheses, as presented in \cref{t:variables}. As discussed in \cref{sss:rq3-analysis}, we decided to use Generalized Additive Models (GAM) to identify the variables that have a significant impact on code reuse.

Since we presumed the variables related to \textbf{H1} and \textbf{H2} would have a linear effect on the probability of code reuse for a project, we kept them as linear terms in the model. The variables related to \textbf{H3} were presumed to have a non-linear effect, so they were used as non-linear terms in the model. The formula we used for invoking the GAM model was:\\
\textit{
Y $\sim$ no.Participant + is.colocated + no.Technology + Before + During + s(pctProse) + s(pctData) + s(pctCode) + s(pctMarkup)}

\begin{table}[t]
\caption{Effect of Project Characteristics on Hackathon Code Reuse - Results from the Generalized Additive Model.\\
\textbf{Part A.} showing the results for the \textit{linear} terms, with the associated Estimate, Standard Error, and  p-Values.\\
\textbf{Part B.} shows the results for the \textit{non-linear} terms, with the Effective Degrees of Freedom -- ``edf'' -- a measure of the degree of non-linearity, the p-Values, and the partial effects of each variable on the response ( 0: No Effect, Positive Values: Positive effects, Negative Values: Negative Effects).\\
The ``pctData'' variable, found to be ``not significant'', is shown in RED, and the corresponding effect plot is omitted}
\label{t:rq3}
\resizebox{\linewidth}{!}{%
\begin{tabular}{@{}p{3cm}llc@{}}
\toprule
\textbf{A. Linear Variables (Hypothesis)} & \textbf{Estimate} & \textbf{Std. Error} & \textbf{p-value} \\ \midrule
no.Participant (H1) & 0.2078 & 0.0181 & $<$ 0.0001 \\
is.colocated-\textbf{TRUE} (H1) & 0.2034 & 0.1030 & 0.0483 \\
no.Technology (H2) & 0.0261 & 0.0060 & $<$ 0.0001 \\
Before (H2) & 0.0001 & 0.0000 & $<$ 0.0001 \\
During (H2) & 0.0036 & 0.0004 & $<$ 0.0001 \\\midrule\midrule
\textbf{B. Non-Linear Variables (Hypothesis) } & \textbf{edf} & \textbf{p-value} & \textbf{Partial Effect Plot} \\\midrule
pctProse (H3) & 3.8984 & 0.0195 & \includegraphics[width=3.5cm]{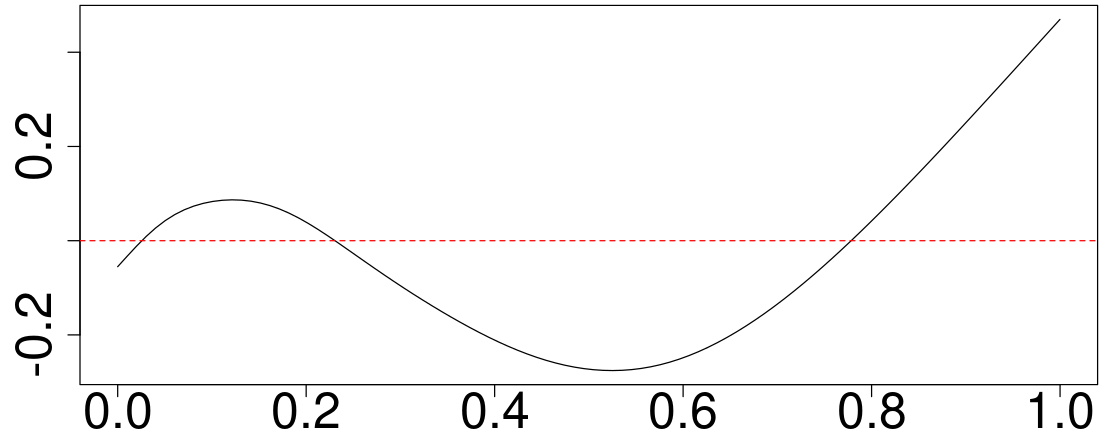} \\\hline
\cellcolor[HTML]{FF8888}pctData (H3) & \cellcolor[HTML]{FF8888}3.7148 & \cellcolor[HTML]{FF8888}0.2490 & \cellcolor[HTML]{FF8888} \\\hline
pctCode (H3) & 3.8425 & 0.0208 &  \includegraphics[width=3.5cm]{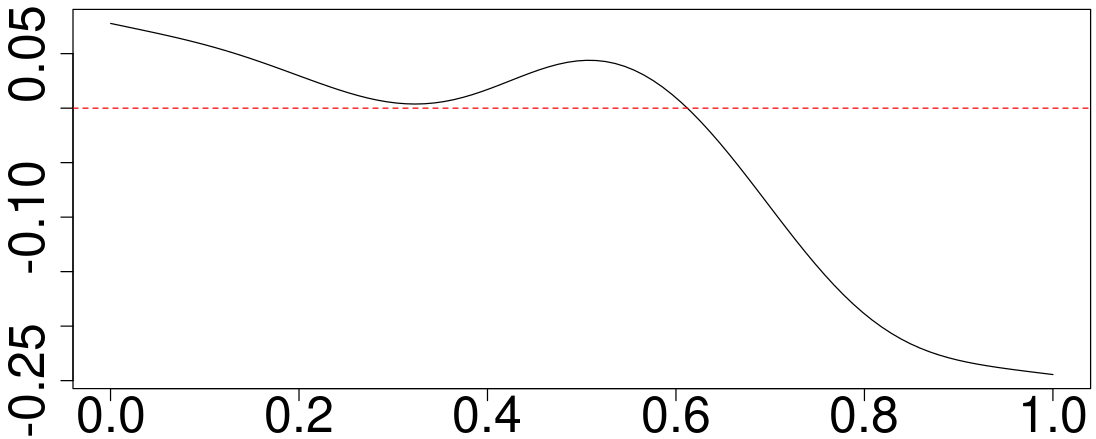} \\\hline
pctMarkup (H3) & 6.6779 & 0.0001 & \includegraphics[width=3.5cm]{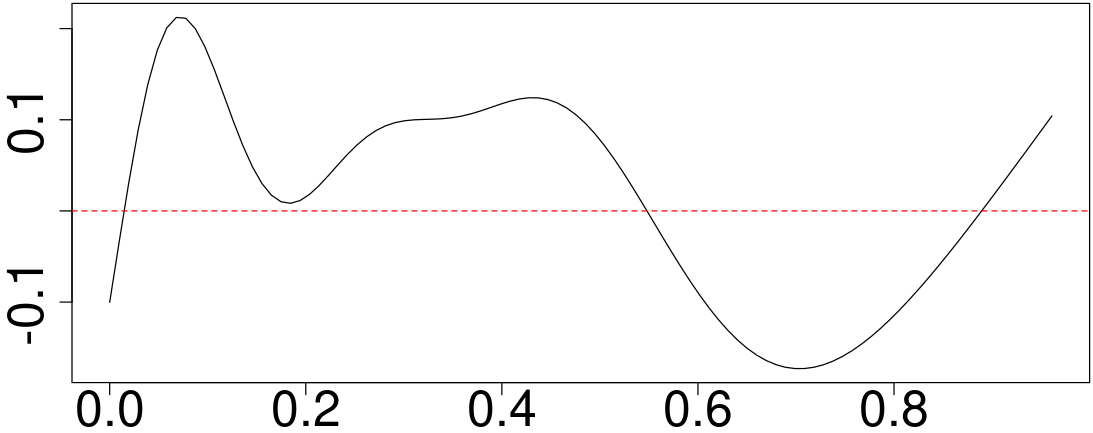} \\ \bottomrule
\end{tabular}%
}
\vspace{-20pt}
\end{table}

The result of the analysis is presented in \cref{t:rq3}, which shows that all of the variables related to hypotheses \textbf{H1} and \textbf{H2}, which we assumed would affect code reuse, were indeed significant. The effect direction, communicated by the signs of the estimates for the corresponding variables match the rationale we presented in \cref{sec:rq}. These findings are partially in line with prior work on hackathon project continuation in that complex projects for which hackathon participants have prepared prior to the hackathon showed increased continuation activity~\cite{nolte2020what}. In that study, the team size was however negatively related to project continuation while for the study in this paper we found the reverse to be true for code reuse. One possible explanation for this discrepancy can be related to larger teams having more opportunities and wider networks to spread the news about their project~\cite{nolte2018you}.

As for the variables related to \textbf{H3}, the ``pctData'' (see \cref{t:variables} for definition) variable was found to be not significant, but the other three variables were. Since the effects of these variables were non-linear in the model, we decided to observe their partial effect plots, which shows the relationship between the two plotted variables (an outcome and an explanatory variable) while adjusting for interference from other explanatory variables~\cite{pplot}. As demonstrated by the partial effect plots, and the effective degrees of freedom associated with each of these variables, each of them actually have a non-linear effect on the response variable. 

Let's take a closer look at the effects of the three significant variables related to \textbf{H3}. For ``pctProse'',  which refers mostly to documentation files (e.g. Markdown, Text, etc.), we see that having some documentation is good, however, projects which have around half its total content as documentations are not likely to have their code reused. However, rather surprisingly, we see that projects with almost all of their content as documentation are more likely to have their code reused. On closer inspection, we found that these are  quite large projects with a lot of data stored in their corresponding repositories in the form of text or other files (An example of such a project is \url{https://github.com/sreejank/PoliClass}). Therefore, though most of them have a good amount of code, by volume, it appears that it is almost all made up of files of type ``Prose'', which causes this predictor to show positive values for projects very high amount of ``Prose'' files. Such projects however are common in particular in civic and scientific hackathons where participants often develop projects that are related to utilizing specific datasets (e.g.~\cite{nolte2020support,trainer2016hackathon,busby2016closing}). Our finding thus can potentially point to a specific use case that is beneficial for code reuse after an event has ended.

On closer inspection of the variable ``pctCode'', we realized that it refers to how much of the total content in a repository is of type ``code'', not the absolute number of code blobs, therefore, having a more balanced repository with a good mix of other types of files signal that it is of higher quality, thus increasing the chance of code reuse. This is somewhat expected since for code to be reused it is beneficial to have accompanying documentation as well as use cases (data). As we can observe, projects with over 60\% of their content related to code take a big hit when it comes to their code getting reused. This finding can potentially be due to hackathon teams only having a finite amount of time during an event to actually develop code and the more the code they develop the more likely it is that they do not have time to ``polish`` it for reuse.

The behavior of the ``pctMarkup'' variable is more complex than the rest (it also has a higher \textit{edf} value), so it is hard to summarize the interaction without a detailed inspection on a larger dataset, however, it looks like having up to around 60\% markup content (e.g. HTML, CSS, LaTeX, etc.) can lead to a higher propensity of code reuse. The reason for the increase for projects with a very high percentage of markup content is likely similar to what we observed for the ``pctProse'' variable.

\hypobox{\textbf{Characteristics affecting Code Reuse (RQ3):} The hypotheses presented in \cref{sec:rq} were found to hold. All of the variables related to \textbf{H1} and \textbf{H2} were significant and had effects as anticipated. The effect of the variables related to \textbf{H3} were more complex, which led to additional insights about hackathon code reuse.}

\vspace{-10pt}
\subsection{A Case-Study on Code Reuse}
\label{sec:results:casestudy}

\begin{table*}[htb]
\caption{Details of the Projects Selected for case study: showing no. of blobs created Before, During, and After the corresponding hackathon events, How many blobs of different types were created \textit{during} the event, and how many of them got reused}
\label{t:case-study}
\resizebox{0.95\linewidth}{!}{%
\begin{tabular}{@{}lllllllllll@{}}
\toprule
\multirow{2}{*}{\textbf{Category}} & \multirow{2}{*}{\textbf{Hackathon Project \textit{(Type)}}} & \multicolumn{3}{c}{\textbf{Blob Creation Time}} & \multirow{2}{*}{\textbf{Blob Reuse}} & \multicolumn{5}{c}{\textbf{Blob Types}} \\
 \cmidrule(lr){3-5} \cmidrule(l){7-11} 
 &  & \textbf{ Before} & \textbf{ During} & \textbf{ After} &  & \textbf{Other} & \textbf{Data} & \textbf{Markup} & \textbf{Code} & \textbf{Prose} \\ \midrule
\multirow{8}{*}{\textbf{\begin{tabular}[c]{@{}l@{}}Projects\\ with\\ code usage\end{tabular}}} 

 & \multirow{2}{*}{Opportunity-Hack-2015-Arizona\_Team1 \textit{(Type-A)} } & \multirow{2}{*}{42} & \multirow{2}{*}{177} & \multirow{2}{*}{\textit{316}} & Reused & 0 & 1 & 0 & 1 & 0 \\ 
 &  &  &  &  & Not reused & 21 & 33 & 19 & 97 & 5 \\ \cmidrule(l){2-11}

& \multirow{2}{*}{TheMichaelHu\_PickyPusheen \textit{(Type-B)}} & \multirow{2}{*}{\textit{3372}} & \multirow{2}{*}{520} & \multirow{2}{*}{1} & Reused & 66 & 3 & 0 & 9 & 1 \\ 
 &  &  &  &  & Not reused & 90 & 259 & 2 & 78 & 12 \\ \cmidrule(l){2-11} 

  & \multirow{2}{*}{drfuzzyness\_WearHax-OlivMatt \textit{(Type-C)}} & \multirow{2}{*}{3154} & \multirow{2}{*}{823} & \multirow{2}{*}{1194} & Reused & 4 & 60 & 0 & 17 & 2 \\ 
 &  &  &  &  & Not reused & 19 & 645 & 0 & 53 & 4 \\ 
  \cmidrule(l){2-11}

 & \multirow{2}{*}{kylemsguy\_building-point \textit{(Type-D)}} & \multirow{2}{*}{29} & \multirow{2}{*}{\textit{81}} & \multirow{2}{*}{10} & Reused & 7 & 7 & 0 & 15 & 0 \\
 &  &  &  &  & Not reused & 0 & 26 & 0 & 26 & 0 \\
 
 \midrule
\multirow{8}{*}{\textbf{\begin{tabular}[c]{@{}l@{}}Projects\\ without\\ code usage\end{tabular}}} 
 
 & \multirow{2}{*}{quki\_IDHACK2016 \textit{(Type-A)}} & \multirow{2}{*}{26} & \multirow{2}{*}{96} & \multirow{2}{*}{\textit{102}} & Used & 0 & 1 & 0 & 0 & 0 \\
 &  &  &  &  & Not reused & 0 & 44 & 0 & 50 & 1 \\ 
\cmidrule(l){2-11} 

 & \multirow{2}{*}{Marblez\_Haven-App \textit{(Type-B)}} & \multirow{2}{*}{\textit{448}} & \multirow{2}{*}{20} & \multirow{2}{*}{1} & Reused & 0 & 0 & 0 & 0 & 0 \\ 
 &  &  &  &  & Not reused & 1 & 11 & 0 & 7 & 1 \\ \cmidrule(l){2-11} 

 & \multirow{2}{*}{shkbfzl\_hs-lunchbot \textit{(Type-C)}} & \multirow{2}{*}{186} & \multirow{2}{*}{48} & \multirow{2}{*}{59} & Reused & 0 & 0 & 0 & 0 & 0 \\ 
 &  &  &  &  & Not reused & 0 & 2 & 0 & 46 & 0 \\ 
 \cmidrule(l){2-11} 

& \multirow{2}{*}{jrdbnntt\_DaReactTV \textit{(Type-D)}} & \multirow{2}{*}{4} & \multirow{2}{*}{\textit{63}} & \multirow{2}{*}{1} & Reused & 2 & 0 & 0 & 0 & 0 \\ 
 &  &  &  &  & Not reused & 2 & 11 & 6 & 41 & 1 \\

 \bottomrule
\end{tabular}
}
\vspace{-15pt}
\end{table*}

In addition to investigating the research questions presented here, we also conducted a small-scale case study on a few projects selected by stratified random sampling to gain additional insights. 

We observed that there are 4 main types of hackathon projects: some containing few blobs that were created before the corresponding hackathon, but have a good amount of activity after the event (Type - \textbf{A }: \textit{After}), while some projects had a large number of blobs that were created before the event, but with little activity afterwards (Type - \textbf{B}: \textit{Before}). Some projects contained code created before, during, and after the event (Type \textbf{C}: \textit{Continuous}), and some mostly contained code created during the event (Type \textbf{D}: \textit{During}).

For this case study, we looked at both projects that had some of their code reused, and projects that did not, and chose one project of each type (A, B, C, and D) for both of these categories at random. The names and details about how many blobs each project had, when they were created, and for the blobs that were created \textit{during} the hackathon, what was their type and how many were reused are presented in \cref{t:case-study}.

Detailed findings for the individual projects with code reuse are listed below:\\
\noindent$\bullet$ \textit{Opportunity-Hack-2015-Arizona\_Team1:} This  project had one code and one data blob reused, both with content related to python libraries, in 2 and 1 other projects respectively.\\
\noindent$\bullet$ \textit{TheMichaelHu\_PickyPusheen}: A number of reuses, mostly icons and configuration from a framework. Only one code blob created by one of the project members was reused, and it was a report for a debug tool they ran.\\
\noindent$\bullet$ \textit{drfuzzyness\_WearHax-OlivMatt:} This project shows evidence of code reuse related to Occulus/Kinect/Wii which was widely reused afterwards by projects of different sizes. All these blobs are under the \texttt{assets} folder and most probably are part of a framework though, so either this team created that framework, or were the first ones to use it. This project had no README or any other documentation.\\
\noindent$\bullet$ \textit{kylemsguy\_building-point:} A good amount of code from this project was used in another small project, likely created by one of the team members. Almost all of the reused files were part of a framework used in the project.

One curious distinction between the projects with and without code reuse was that, in most cases, projects with code reuse had other types of blobs reused as well, and for those without code reuse, they mostly have none of their blobs reused. This is in line with the previously discussed finding that additional documentation and a potential use case can foster code reuse. It is also worth noting that for almost all of the code reuses we observed, the code was part of a package/library/framework used in the project. This is not too surprising, since we are only looking at \textit{exact} code reuse. However, we filtered our dataset to only look for blobs first created by one of the members of the project team during the event, so it is very likely that the framework was created/modified to some extent by the project team members, and that newly created/modified framework was later used by others. At the same time, it is possible that someone might have made the exact same changes to a file as made by a member of the hackathon project, which then gets counted as reuse. Further study is needed to ascertain the probability of such chance events happening by accident.

\section{Implications}
\label{sec:impl}
Our findings have a number of implications for research and practice. They can serve as a valuable guidance for scientific and other communities that aim to organize hackathons for expanding their existing code base. Organizers could suggest participating teams to attempt projects that do not require  developing a large amount of code and rather focus on a specific use case e.g. related to an existing data set. Moreover, they should suggest teams to also spend time on not only developing code but also providing additional materials and documentation, and also for the teams to reuse existing code from their existing code base rather than attempting to develop a lot of original code. This approach can in turn improve efficiency and foster code reuse after an event has ended.

With respect to research, our findings provide an initial account on how code gets reused and created during a hackathon as well as whether and where it gets reused afterwards. They indicate that not much new code gets developed during a hackathon and much of the code used by the teams is actually reused from existing code, thus altering our perception that hackathons are intensive code creation events. Moreover, they indicate that hackathon code indeed gets reused and that hackathons can thus be more than one-off coding events.

\vspace{-5pt}
\section{Limitations and Threats to Validity}
\label{s:limitation}
For our study, we tracked the code generation and usage on a blob level - represented in \WOC by the SHA1 hash value of each blob - which means that we focused only on exact code reuse since any changes in the file contents would lead to a change in the blob SHA1 value that we used to identify each blob. However, it is quite common to make minor changes in a code file while using it in a different context, and that aspect will not be captured in our study, nor can we capture the reuse of code snippets. Moreover, our analysis does not consider the size of a file since we aree looking at the SHA1 values of the blobs, i.e. our analysis cannot distinguish between reuse of a small file and that of a large file. 

The \DP dataset does not include the start date of the hackathon events but it is essential information needed to answer our research questions. We assumed the duration of the hackathons to be 72 hours based on existing literature and a manual investigation of 73 randomly selected hackathons 71 of which lasted up to 3 days. However, that might not have been the case for all of the events we studied which may affect the results of RQ1 and RQ3.

We relied on the \GH Linguist tool to categorize files and we only focused on files with type "Programming", however the categorization is not infallible, e.g. the type "Markup" contains HTML and CSS files which could be considered code instead of documentation. 

Finally, in our study we only considered hackathon projects, thus, our findings may not be generalizable to other types of software projects and repositories.




\vspace{-5pt}
\section{Conclusion and Future Work}
In this study, we investigated the origins of hackathon code and its reuse after an event. We found that most hackathon projects reuse existing code and that code created during events also gets reused by other OSS projects later on. Our study also revealed that project characteristics related to its prolificness and the developers' familiarity with the code positively affects code reuse, and the composition of the project in terms of what file types it contains have an effect as well. In summary, our findings agree with most earlier studies, and reiterate the impact of hackathon events, at the same providing an account of code reuse in Open Source Software.

There are several ways to extend this research, e.g. considering code clones/snippets while looking for code reuse (e.g. by looking at the associated CTAG tokens - a dataset available in \WOC), identifying other factors that affect code reuse, including code quality~\cite{dey2018usageQuality,Dey2020qualityEMSE}, project popularity~\cite{dey2018dependency,dey2019patterns}, the type of Open Source license used, etc. Looking deeper into the code created during the hackathons, it might also be interesting to see to what extent the teams use bots~\cite{dey2020botdetection,dey2020botse} which might aid in the understanding of hackathon code reuse as well.
We hope that further studies will explore these and other related topics, and give us a clearer understanding of the impact of hackathons and code reuse.


\vspace{-10pt}
\section*{Acknowledgments}
The work was supported, in part, by Science Foundation Ireland grant 13/RC/2094\_P2 and by NSF awards 1633437, 1901102, and 1925615.



\bibliographystyle{IEEEtran}
\bibliography{references}

\end{document}